\documentclass[useAMS,usenatbib]{mnras}
\usepackage{longtable}
\usepackage{graphicx}
\usepackage{lscape}
\usepackage{rotating}

\usepackage[T1]{fontenc}
\usepackage{ae,aecompl}

\newcommand{\Msun}{\mbox{$\mathrm{M}_{\odot}$}}

\newcommand{\Teff}{\mbox{$T_{\mathrm{eff}}$}}

\newcommand{\Ion}[2]{#1{\,\scriptsize #2}}

\newcommand{\kms}{\mbox{$\mathrm{km\,s^{-1}}$}}

\title[The white dwarf binary pathways survey] {The white dwarf binary
  pathways survey - II. Radial velocities of 1453 FGK stars with white
  dwarf companions from LAMOST DR\,4}

\author[A.  Rebassa-Mansergas et al.]{A.  Rebassa-Mansergas$^{1,2}$,
  J.J.  Ren$^{3}$, P.  Irawati$^{4}$, E. Garc\'ia-Berro$^{1,2}$,
  S.G. Parsons$^{5}$, \newauthor M.R.  Schreiber$^{6,7}$, B.T.
  G\"ansicke$^{8}$, P.  Rodr\'iguez-Gil$^{9,10}$, X.  Liu$^{11}$, C.
  Manser$^{8}$, \newauthor S.  Palomo-Nevado$^{9,10}$, F.
  Jim\'enez-Ibarra$^{9,10}$, R.  Costero$^{12}$, J.
  Echevarr\'ia$^{12}$, \newauthor R. Michel$^{13}$, M.
  Zorotovic$^{6}$, M.  Hollands$^{8}$, Z. Han$^{14}$, A. Luo$^{3}$,
  E. Villaver$^{15}$, Kong, X.$^{3}$\\
$^{1}$ Departament de F\'\i sica, Universitat Polit\`ecnica de
  Catalunya, c/Esteve Terrades 5, 08860 Castelldefels, Spain\\
$^{2}$ Institute for Space Studies of Catalonia, c/Gran Capit\`a 2--4,
  Edif. Nexus 201, 08034 Barcelona, Spain\\
$^{3}$ National Astronomical Observatories, Chinese Academy of
  Sciences, 100012 Beijing, P.\,R.\,China\\
$^{4}$ National Astronomical Research Institute of Thailand, 191
  Siriphanich Building, Huay Kaew Road, Chiang Mai 50200, Thailand\\
$^{5}$  Deparment of  Physics  \& Astronomy,  University of  Sheffield,
  Sheffield S3 7RH, UK\\
$^{6}$  Instituto  de F\'\i  sica  y  Astronom\'\i a,  Universidad  de
 Valpara\'\i so, Avenida Gran Breta\~na 1111, Valpara\'\i so, Chile \\
$^{7}$ Millenium Nucleus "Protoplanetary Disks in ALMA Early Science",
 Universidad  de   Valpara\'\i  so,   Avenida  Gran   Breta\~na  1111,
 Valpara\'\i so, Chile\\
$^{8}$ Department of Physics, University of Warwick, Coventry CV4 7AL,
 UK \\
$^{9}$ Instituto de Astrof\'isica de Canarias, c/ V\'ia L\'actea s/n,
 E-38205 La Laguna, Tenerife, Spain\\
$^{10}$ Universidad de La Laguna, Departamento de Astrof\'isica, E-
 38206 La Laguna, Tenerife, Spain\\
$^{11}$ Department of Astronomy, Peking University, Beijing 100871,
 P.\,R.\,China\\
$^{12}$ Instituto de Astronom\'ia, Universidad Nacional Aut\'onoma de
 M\'exico, 70-264, Ciudad Universitaria, M\'exico
 D.F., C.P. 04510, M\'exico\\
$^{13}$ Instituto de Astronom\'ia, Universidad Nacional Aut\'onoma de
 M\'exico, 877, Ensenada, Baja California, C.P. 22830
 M\'exico\\
$^{14}$ Key Laboratory for the Structure and Evolution of Celestial
 Objects, Yunnan observatories, Chinese Academy of Sciences, P.O. Box
 110, \\ Kunming, 650011, Yunnan Province, P.\,R.\,China\\
$^{15}$ Departamento de F\'isica Te\'orica, Universidad Aut\'onoma de
 Madrid, Cantoblanco 28049 Madrid, Spain
}

\begin{document}
\date{Accepted 2017. Received 2017; in original form 2017}
\pagerange{\pageref{firstpage}--\pageref{lastpage}}
\pubyear{2016}
\maketitle

\begin{abstract}
We present  the second  paper of  a series  of publications  aiming at
obtaining  a better  understanding  regarding the  nature  of type  Ia
supernovae (SN\,Ia) progenitors by studying a large sample of detached
F,  G and  K main  sequence  stars in  close orbits  with white  dwarf
companions (i.e.  WD+FGK  binaries).  We employ the  LAMOST (Large Sky
Area  Multi-Object  Fibre  Spectroscopic  Telescope)  data  release  4
spectroscopic  data  base  together  with  $GALEX$  (Galaxy  Evolution
Explorer)  ultraviolet   fluxes  to   identify  1,549   WD+FGK  binary
candidates (1,057 of which are new), thus doubling the number of known
sources. We measure  the radial velocities of 1,453  of these binaries
from the available  LAMOST spectra and/or from spectra  obtained by us
at  a wide  variety  of  different telescopes  around  the globe.  The
analysis of the radial velocity data  allows us to identify 24 systems
displaying  more  than 3$\sigma$  radial  velocity  variation that  we
classify  as close  binaries. We  also discuss  the fraction  of close
binaries among WD+FGK systems, which we  find to be $\sim$10 per cent,
and  demonstrate  that  high-resolution spectroscopy  is  required  to
efficiently identify double-degenerate SN\,Ia progenitor candidates.
\end{abstract}


\label{firstpage}

\section{Introduction}
\label{s-intro}

Thermonuclear supernovae,  also known  as Type Ia  supernovae (SN~Ia),
originate   from  the   explosion   of   carbon-oxygen  white   dwarfs
\citep[WDs;][]{HoyleFowler}, and are among the most luminous events in
the Universe.   Consequently, SN~Ia can  be detected up to  very large
distances.  Since  they arise from the  detonation of WDs in  a narrow
mass range, they  have nearly the same intrinsic  luminosity, and thus
they play  a primary role  as standardizable candles in  cosmology. In
particular,  the  discovery  of  a relation  between  their  intrinsic
luminosity and  the shape of  their light curves  \citep{Phillips} has
paved the  way to  a new  era of precision  cosmology, leading  to the
discovery of  the acceleration  of the  Universe \citep{riessetal98-1,
  Schmidtetal98, Perlmutteretal99, astier+pain12-1}. However, although
SN~Ia observations have led to  fundamental discoveries, the nature of
their  progenitors is  not yet  unambiguously determined.   Even more,
albeit their  intrinsic importance,  the explosion mechanism  of SN~Ia
still  remains  poorly understood  after  more  than five  decades  of
theoretical and  observational work,  thus compromising their  use and
possibly introducing some not yet  well-known systematic errors in the
determination   of   extragalactic  distances   \citep{lindenetal09-1,
  howell11-1}.

It is widely accepted that a WD in  a binary system is at the heart of
a SN~Ia  outburst. Several  possible evolutionary channels  leading to
such an  event are currently  envisaged, all of which  have advantages
and drawbacks.   Hence, there is  not yet  a general consensus  on the
leading path for SN~Ia. In fact, it  is well possible that all of them
contribute   to   the   total    rate   in   some   unknown   fraction
\citep{wang+han12-1}.    These  evolutionary   channels  are   briefly
described below.

In the  single degenerate channel  \citep{whelan+iben73-1, nomoto82-1,
  han+podsiadlowski04-1} the  WD accretes  mass from  a non-degenerate
stellar donor and explodes when  its mass grows near the Chandrasekhar
limit.  In the  double-detonation mechanism \citep{woosley+weaver94-1,
  livne+arnett95-1} a  WD with a  mass smaller than  the Chandrasekhar
limit  accumulates  helium-rich  material  on  its  surface  from  its
companion.   The  helium  layer  is compressed  as  more  material  is
accreted  and ultimately  detonates. The  compression wave  propagates
towards the center  and a second detonation occurs near  the center of
the carbon-oxygen  core -- see, for  instance, \cite{shenetal12-1} and
references   therein.   Next   comes  the   double-degenerate  channel
\citep{whelan+iben73-1,   nomoto82-1,   iben+tutukov84-1}.   In   this
scenario two  WDs merge and  a prompt, violent detonation  ensues when
the masses of the two merging  WDs are sufficiently large, otherwise a
delayed explosion could eventually  occur.  Recently, another possible
scenario  has been  proposed,  the  so-called core-degenerate  channel
\citep{livio+riess03-1,  kashi+soker11-1, sokeretal13-1},  where a  WD
merges with  the hot core of  a massive asymptotic giant  branch (AGB)
star  during  or after  a  common  envelope phase.   Finally,  another
possible channel involves the collision of  two WDs in a dense stellar
environment  \citep{thomsonetal11-1,  katz+dong11-1,  kushniretal13-1,
  aznar-siguanetal13-1}.   In this  scenario  either  a tertiary  star
brings two  WDs to collide, or  the dynamical interaction occurs  in a
dense stellar system, where such  interactions are likely. It has been
shown  that, in  some cases,  the  collision results  in an  immediate
explosion. However, this  scenario can only account for  a few percent
of    all    SN~Ia    under   the    most    favourable    assumptions
\citep{hamerseta13-1,  sokeretal14-1}.  In  summary, the  challenging,
unsolved problem of understanding  and quantifying the contribution of
the different progenitor channels to  the observed population of SN~Ia
is still a matter of active research.

Currently, the  two most popular  scenarios for SN~Ia are  the single-
and the  double-degenerate channels. The  viability of both  paths has
been  intensively  studied  during  the  last  several  years  from  a
theoretical perspective \citep[e.g.][]{hachisuetal12-1, boursetal13-1,
  nielsenetal14-1,         jimenezetal16-1,         vanrossumetal16-1,
  yungelson+kuranov17-1}.   Unfortunately,  these theoretical  studies
have  yielded so  far  no conclusive  results on  which  of these  two
channels is  more efficient  in producing  SN\,Iae. For  instance, the
mass growth of  the WD in the single-degenerate channel  is thought to
be rather small \citep[e.g.][]{idanetal12-1}  avoiding reaching a mass
near the  Chandrasekhar limit.  Moreover, determinations  of the SN~Ia
delay  time distributions  are  not consistent  with expectations  for
single-degenerate systems \citep{galyam+maoz04-1, totanietal08-1}.  In
contrast,  the   double-degenerate  channel  predicts  a   delay  time
distribution     in     better     agreement     with     observations
\citep{maoz+badenes10-1, maozetal10-1}.   However, current simulations
predict that the range of  WD masses that produce powerful detonations
is rather limited, and in most of  the cases the outcome of the merger
could   result  in   other   (interesting)  astrophysical   phenomena.
Specifically, it  has been found that  a possible outcome could  be an
accretion-induced collapse  to a neutron  star \citep{nomoto+iben85-1,
  shenetal12-1}, whereas in some other  cases a high-field magnetic WD
could   be   formed  \citep{garcia-berroetal12-1}.    Observationally,
several additional  analyses have provided  some support for  both the
single-   \citep{livio+riess03-1,  hamuyetal03-1,   voss+nelemans08-1,
  shappeeetal16-1,   liu+stancliffe16-1}  and   the  double-degenerate
channels     \citep{gonzalez-hernandezetal12-1,     santanderetal15-1,
  ollingetal15-1}. However,  there is  no single  system yet  that has
robustly  been  confirmed as  a  single-  or double-degenerate  SN\,Ia
progenitor -- see, e.g., \cite{garcia-berroetal16-1}.

With these thoughts in mind  we have started an observational campaign
that aims at providing additional observational input for testing both
the  single-  and  double-degenerate  channels  for  SN\,Ia  based  on
observations of  WDs in  detached close  binaries where  the secondary
star is an  F, G or K main sequence  star (hereafter WD+FGK binaries).
These  systems  were initially  main  sequence  binaries that  evolved
through common envelope evolution and have now orbital periods ranging
from a few hours to several weeks \citep{willems+kolb04-1}.  Depending
on the orbital  periods and component masses, the  WD+FGK binaries are
expected to evolve  mainly through the following two  channels. In the
first path  the secondary star transfers  mass to the WD  in a thermal
timescale,  ensuing the  so-called super-soft  (SSS) phase.   The mass
transfer rate is high enough to sustain stable hydrogen burning so the
WD   mass   grows.    Hence,    the   system   becomes   a   potential
single-degenerate  SN\,Ia  progenitor \citep{sharaetal77-1,  iben82-1,
  fujimoto82-1,  vandenheuveletal92-1,  distefano10-1,  wolfetal13-1}.
If the  WD does not accrete  sufficient mass during the  SSS phase the
WD+FGK becomes a cataclysmic variable with a massive WD and an evolved
donor       \citep[e.g.][]{schenkeretal02-1,      thorstensenetal02-2,
  gaensickeetal03-1,                                zorotovicetal11-1,
  rebassa-mansergasetal14-1}. Another  alternative is that  the system
may go  through a  second common  envelope phase.   In this  case, the
envelope may  be ejected  and the  result is  a double  WD in  a tight
orbit,   i.e.   a   potential  double-degenerate   SN\,Ia  progenitor.
Depending on the time that it takes  to bring the two WDs close enough
to overflow the Roche lobe (it should be shorter than the Hubble time)
and  the  total  mass  of  the  system (it  should  be  close  to  the
Chandrasekhar mass) the result could be a SN~Ia outburst. By analysing
a large sample  of close WD+FGK binaries with  well determined orbital
periods and component masses we will be able to reconstruct their past
evolution, and to predict their future evolution and thus quantify the
fraction of  systems evolving through  each path.  In is  important to
keep in  mind that our  aim here is  not to identify  individual SN~Ia
progenitors (in  fact the vast  majority of our close  WD+FGK binaries
will not  become SN~Ia), but to  test our understanding of  the binary
evolution theory of the single- and double-degenerate channels towards
SN~Ia.

\begin{figure}
  \begin{center}
    \includegraphics[angle=-90, width=\columnwidth]{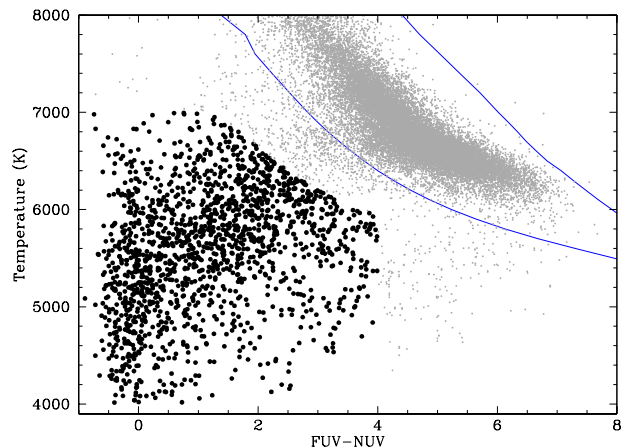}
    \caption{\label{f-selection}  The UV  colours and  temperatures of
      LAMOST main-sequence FGK stars (gray dots). The temperatures are
      taken directly from  the LAMOST DR\,4 catalogue.   Shown in blue
      are  two theoretical  tracks computed  from PHOENIX  atmospheric
      models, one  with a  high metallicity  and high  surface gravity
      ($\log$\,Z = +1  and $\log$\,g = 5.0, top curve)  and one with a
      low  metallicity and  low surface  gravity ($\log$\,Z  = -3  and
      $\log$\,g = 3.5, bottom  curve).  These represent extreme limits
      on  the colours  of main-sequence  stars. The  vast majority  of
      stars fall within these extreme  models, as expected.  The black
      solid points mark the 1,599 targets that fall within our cut for
      selecting WD+FGK  binary candidates  (FUV-NUV $<$  4, 8,000$>T>$
      4,000\,K, and  FUV-NUV at  least 1.5  magnitudes bluer  than the
      bottom PHOENIX model).}
  \end{center}
\end{figure}

In  \citet{parsonsetal16-1} we  presented  the first  of  a series  of
publications resulting from our dedicated campaign, where we described
our  methodology   to  efficiently   identify  WD+FGK   binaries  (see
\citealt{parsonsetal15-1} for  our study  on the first  SSS progenitor
identified  so far).   These  were  selected as  F,  G,  K stars  with
northern-hemisphere  LAMOST \citep[Large  Sky Area  Multi-Object Fibre
  Spectroscopic     Telescope;][]{cuietal12-1,    zhaoetal12-1}     or
southern-hemiesphere        RAVE         \citep[Radial        Velocity
  Experiment;][]{kordopatisetal13-1}  spectra  displaying  ultraviolet
(UV)  excess as  indicated  by their  $GALEX$ \citep[Galaxy  Evolution
  Explorer;][]{martinetal05-1} FUV-NUV colours (where  FUV and NUV are
the far-UV and  near-UV magnitudes) in a \Teff\,  vs.  FUV-NUV diagram
(see Figure\,\ref{f-selection}).   This resulted in 430  WD+FGK binary
candidates from RAVE DR\,1 and 504 candidates from LAMOST DR\,1.

In this paper we first expand our WD+FGK binary sample by applying our
selection criteria  to the  FGK star catalogue  from LAMOST  DR\,4. We
then  use LAMOST  and  follow-up spectroscopic  observations of  1,453
WD+FGK binary candidates in the  new LAMOST sample for measuring their
radial  velocities  and  we  identify  24  close  binaries  displaying
significant radial velocity variations.  Finally, we discuss the close
binary fraction  among WD+FGK binaries.  In  a forthcoming publication
we will present a radial  velocity analysis dedicated to follow-up our
southern-hemisphere RAVE targets  as well as our  first orbital period
measurements.

\section{The DR\,4 of LAMOST}

LAMOST  is  a  $\simeq$\,4\,meter  quasi-meridian  reflecting  Schmidt
telescope located at Xinglong Observing  Station in the Hebei province
of China \citep{cuietal12-1}.  It has a  field of view of 5$^\circ$ in
diameter   and  it   is  exclusively   dedicated  to   obtain  optical
spectroscopy  of celestial  objects.  Each  ``spectral plate''  refers
physically to a focal surface  with 4,000 precisely positioned optical
fibers   to    observe   spectroscopic   plus    calibration   targets
simultaneously, equally distributed  among 16 fiber-fed spectrographs.
Each spectrograph  is equipped with  two CCD  cameras of blue  and red
channels that simultaneously provide blue and red spectra of the 4,000
selected    targets,   respectively.     The    spectra   cover    the
$\sim$3,700-9,000\AA\,  wavelength  range  at  a  resolving  power  of
$\sim$1,800.

Since September  2012 LAMOST has  been performing a  five-year regular
survey, which  was preceded by  a two-year commissioning survey  and a
one-year pilot survey. The LAMOST  regular survey consists of two main
parts \citep{zhaoetal12-1}:  the LAMOST Extra-Galactic  Survey (LEGAS)
of galaxies  to study the large  scale structure of the  Universe, and
the  LAMOST  Experiment  for Galactic  Understanding  and  Exploration
(LEGUE) Survey of the Milky Way  which is developed to obtain millions
of stellar spectra to study the  structure and evolution of the Galaxy
\citep{dengetal12-1}.   LEGUE is  sub-divided  into three  independent
surveys for follow-up observations of  the spheroid, the disk, and the
galactic     anti-centre,      respectively     \citep{carlinetal12-1,
  chenetal12-1, yuanetal15-2, xiangetal17-1, xiangetal17-2}.

\begin{table*}
\centering
\caption{\label{tab:wdfgk-sample} The selected  2,168 spectra of 1,549
  unique    LAMOST    DR\,4     WD+FGK    binary    candidates    (see
  Figure\,\ref{f-selection}),   including   coordinates,  $GALEX$   UV
  magnitudes, spectral ID identifiers  (plate, spectrograph, fibre IDs
  and modified  Julian date  MJD), signal-to-noise ratio  in different
  spectral bands  and stellar parameters.  The  ``Mag" column provides
  the  optical  magnitudes obtained  from  AAVSO  Photometric All  Sky
  Survey DR9  \citep[APASS;][]{hendenetal16-1}, although  some entries
  are      obtained     from      the     Catalina      Sky     Survey
  \citep[CSS;][]{drakeetal09-1} or  Guiding Star Catalog  (GSC).  Only
  for few systems  with no available $V$ magnitudes we  report the $g$
  magitude     from      the     Sloan     Digital      Sky     Survey
  \citep[SDSS;][]{yorketal00-1}   or   the  Xuyi   Schmidt   Telescope
  Photometric     Survey      of     the      Galactic     Anti-center
  \citep[XSTPS-GAS;][]{liuetal14-1,   zhangetal14-1}.   The   complete
  table  can  be  found  in  the electronic  version  of  the  paper.}
\setlength{\tabcolsep}{2pt} \resizebox{\textwidth}{!}{%
\begin{tabular}{lrrrrrrrrcccccrrrrrrrrrrrc}
  \hline
                 Jname  &       RA      &       DEC     &       FUV     &       Err & NUV & Err & Mag & Err & Mag & MJD & Plateid & Spid & Fiberid\\
                 &  ($^\circ$) &   ($^\circ$) & (mag) & (mag) & (mag) & (mag) &  (mag) & (mag) & from & & & & \\
\hline
J000230.57+155618.1     &       0.6273770       &       15.9383670      &       18.184  &       0.060   &       17.632  &       0.025   &       13.084  &       0.015   &       APASS V &       56199   &       EG000313N173308V$\_$1   &       01      &       151     \\
J000232.64+074305.5     &       0.6360292       &       7.7182140       &       17.369  &       0.045   &       17.864  &       0.034   &       16.770  &       0.080   &       CSS V   &       57309   &       EG000015N080026M01      &       04      &       110     \\
J000243.58+270214.7     &       0.6816070       &       27.0374370      &       19.704  &       0.121   &       19.564  &       0.073   &       14.660  &       0.026   &       APASS V &       56648   &       VB003N26V2              &       10      &       134     \\
J000324.28+063732.3     &       0.8512020       &       6.6256600       &       19.531  &       0.141   &       17.034  &       0.027   &       12.861  &       0.022   &       APASS V &       56618   &       EG000954N044957V01      &       16      &       031     \\
J000545.12+410828.0     &       1.4380150       &       41.1411130      &       20.647  &       0.197   &       18.100  &       0.019   &       13.524  &       0.048   &       APASS V &       56611   &       VB002N40V3              &       15      &       229     \\
J000640.63+024704.9     &       1.6693030       &       2.7847050       &       20.638  &       0.184   &       18.941  &       0.052   &       15.269  &       0.093   &       APASS V &       55893   &       F9302                   &       02      &       125     \\
J000652.08+012950.4     &       1.7170400       &       1.4973470       &       19.864  &       0.082   &       16.831  &       0.013   &       11.200  &       0.171   &       APASS V &       56230   &       EG001639N015102V01      &       10      &       127     \\
J000719.16+261658.2     &       1.8298700       &       26.2828580      &       19.976  &       0.113   &       17.511  &       0.014   &       11.428  &       0.092   &       APASS V &       56648   &       VB003N26V1              &       10      &       201     \\
J000752.32+083234.1     &       1.9680120       &       8.5428300       &       18.601  &       0.089   &       18.523  &       0.058   &       14.932  &       0.018   &       APASS V &       56919   &       EG001605N080655B01      &       14      &       226     \\
J000920.63+071827.7     &       2.3359708       &       7.3077083       &       20.735  &       0.150   &       20.581  &       0.101   &       14.744  &       0.086   &       APASS V &       56618   &       EG000954N044957B01      &       11      &       149     \\
J000920.63+071827.7     &       2.3359708       &       7.3077083       &       20.735  &       0.150   &       20.581  &       0.101   &       14.744  &       0.086   &       APASS V &       56919   &       EG001605N080655B01      &       10      &       029     \\
\end{tabular}}
\setlength{\tabcolsep}{6pt}
\resizebox{\textwidth}{!}{%
\begin{tabular}{lrrrrrrrrcccccrrrrrrrrrrr}
  \hline
                 Jname & S/N$_u$ & S/N$_g$ & S/N$_r$ & S/N$_i$ & S/N$_z$ & \Teff & Err & $\log$\,g  & Err & [Fe/H] & Err \\
                   & & & & & & (K) & (K) & (dex) & (dex) & (dex) & (dex)  \\
\hline
J000230.57+155618.1     &       8.77    &       22.23   &       17.70   &       6.69    &       6.43    &       5982.68 &       369.96  &       4.165   &       0.607   &       0.144   &       0.318     \\
J000232.64+074305.5     &       4.01    &       13.27   &       18.65   &       23.53   &       12.87   &       5419.72 &       309.29  &       4.493   &       0.444   &       -0.726  &       0.287     \\
J000243.58+270214.7     &       4.96    &       29.53   &       47.19   &       59.20   &       36.35   &       5488.39 &       123.38  &       4.550   &       0.176   &       0.123   &       0.115     \\
J000324.28+063732.3     &       36.03   &       118.83  &       147.94  &       164.79  &       114.54  &       6253.46 &       9.89    &       4.007   &       0.013   &       -0.377  &       0.009     \\
J000545.12+410828.0     &       25.94   &       69.90   &       62.03   &       39.39   &       29.83   &       6317.23 &       195.71  &       4.248   &       0.464   &       0.065   &       0.176     \\
J000640.63+024704.9     &       6.37    &       32.87   &       40.40   &       45.46   &       26.42   &       6419.92 &       161.80  &       4.152   &       0.231   &       -0.349  &       0.151     \\
J000652.08+012950.4     &       10.11   &       43.87   &       23.26   &       21.62   &       21.37   &       5787.07 &       45.47   &       4.401   &       0.064   &       0.348   &       0.043     \\
J000719.16+261658.2     &       52.35   &       171.30  &       272.97  &       330.85  &       245.20  &       5295.79 &       5.64    &       4.412   &       0.005   &       -0.035  &       0.005     \\
J000752.32+083234.1     &       8.85    &       68.12   &       118.39  &       162.01  &       113.39  &       5534.04 &       39.37   &       4.225   &       0.055   &       -0.216  &       0.037     \\
J000920.63+071827.7     &       8.13    &       60.83   &       116.79  &       155.20  &       109.27  &       5240.46 &       43.34   &       4.502   &       0.061   &       -0.424  &       0.041     \\
J000920.63+071827.7     &       9.12    &       57.54   &       97.22   &       120.20  &       74.40   &       5243.20 &       48.97   &       4.514   &       0.069   &       -0.437  &       0.046     \\
\hline
\end{tabular}}
\end{table*}

The  raw  spectra  are  reduced by  the  LAMOST  two-dimensional  (2D)
pipeline \citep{luoetal12-1, luoetal15-1}, which extracts spectra from
raw CCD images  and calibrates them. This procedure  includes dark and
bias subtraction,  cosmic ray  removal, one-dimensional  (1D) spectral
extraction,   flat-field  correction,   wavelength  calibration,   sky
subtraction,  merging  sub-exposures  and combining  wavelength  bands
(after  flux calibration).   Each  LAMOST spectrum  is  the result  of
combining a certain number  of individual sub-exposures ($\geq$3), and
in several cases  the targets are observed in different  nights.  In a
second  step the  1D pipeline  works on  spectral type  classification
(four  primary  types:  star,  galaxy,  quasar,  unknown)  and  radial
velocity (or redshift) measurement for galaxies and quasars.  Finally,
the   LAMOST  Stellar   Parameter  Pipeline   (LASP)  accurately   and
automatically determines the fundamental stellar parameters for late A
and FGK  type stellar spectra  (effective temperature \Teff  , surface
gravity: $\log$\,g , metallicity [Fe/H],  and radial velocity RV).  To
that  end LASP  adopts two  consecutively methods  -- the  Correlation
Function       Initial        (CFI)       value        and       UlySS
\citep[Universit$\acute{\mathrm{e}}$  de  Lyon Spectroscopic  analysis
Software;][]{wuetal11-1}.  The CFI method is  used to produce a set of
coarse measurements  that serve  as initial  guesses for  ULySS, which
then determines the  final stellar parameters.  By  comparing with the
estimates from  high resolution spectra,  the \Teff , $\log$\,g  , and
[Fe/H]  determined by  LASP  have the  external errors  $-47\pm95$\,K,
$0.03\pm0.25$\,dex,      and      $-0.02\pm0.1$\,dex      respectively
\citep{luoetal15-1}.

The latest data product of LAMOST is Date Release 4 (DR\,4), currently
only    released    internally    to    the    Chinese    astronomical
community\footnote{http://dr4.lamost.org/}.   DR\,4  includes  spectra
observed from October 24th, 2011 to  June 2nd, 2016 by 3,454 different
plates.  DR\,4 thus contains a  total of 7\,681\,185 spectra, of which
6\,898\,298 are catalogued as stars,  118\,743 as galaxies, 41\,352 as
quasars and 622\,792 are unknown objects.  Furthermore, in addition to
the spectral  data and the  general catalogue mentioned  above, LAMOST
DR\,4  also  contains  three  catalogues  including  measured  stellar
parameters for (1)  late A and FGK-type main sequence  stars with high
quality spectra (4\,202\,127 entries),  (2) A-type main sequence stars
(364\,600  entries),  and (3)  M-type  main  sequence stars  (433\,247
entries).

\section{WD+FGK binaries in LAMOST DR\,4}
\label{s-selection}

WD+FGK binaries are not easy to identify in the optical since the main
sequence       stars       generally      outshine       the       WDs
\citep{rebassa-mansergasetal10-1}.  Hence, UV  coverage is required to
detect    the    WD.    As    we    have    already   mentioned,    in
\citet{parsonsetal16-1}   we  developed   a  method   for  efficiently
identifying  WD+FGK binaries  based  on the  following criteria  (also
illustrated in Figure\,\ref{f-selection}):

\begin{eqnarray}
eFUV \leq 0.2,   eNUV \leq 0.2,   (FUV-NUV) \leq 4, \\
4000\,\mathrm{K} \leq \Teff \leq 7000\,\mathrm{K}, e\Teff \leq 
500\,\mathrm{K}, \\
\Teff<8592.91552-2747.51667\times(FUV-NUV)+ \nonumber\\
+2292.51782\times(FUV-NUV)^2-1279.40096\times \nonumber\\
\times(FUV-NUV)^3+407.13454\times \nonumber\\
\times(FUV-NUV)^4-71.88852\times \nonumber\\
\times(FUV-NUV)^5+6.56399\times \nonumber\\
\times(FUV-NUV)^6-0.24160\times(FUV-NUV)^7\\
NUV\_art=0, FUV\_art=0,
\end{eqnarray}

\noindent where  $eFUV$ and $eNUV$  are the $NUV$ and  $FUV$ magnitude
errors, \Teff\,  is the effective  temperature provided by  the LAMOST
DR\,4 stellar parameter pipeline and NUV$\_$art and FUV$\_$art are the
NUV  and  FUV  artifacts,  i.e.  we  choose  only  $GALEX$  photometry
associated to  no artifacts.  This  results in 2,226 spectra  of 1,599
unique  main  sequence  FGK  stars being  selected  as  WD+FGK  binary
candidates (black  solid dots in Figure\,\ref{f-selection})  among the
4\,202\,127 entries in the LAMOST DR\,4 late A, FGK-type main sequence
stellar parameter  catalogue. Visual inspection of  the 2,226 selected
spectra revealed the typical features of FGK stars in all cases except
for seven objects associated to bad quality spectra that we decided to
exclude.   Moreover, we  cross-correlated our  list with  the $Simbad$
catalogue  and found:  nine  eclipsing  binaries of  W  UMa type,  one
eclipsing main sequence binary, one  eclipsing binary of Lyr type, one
eclipsing binary  of Algol type,  two classical cepheids,  26 variable
stars of RR  Lyr type, one symbiotic star, one  red giant branch star,
one T  Tauri star and  three RS CVn  binaries.  We excluded  all these
sources  from our  list  except the  three RS  CVn  binaries since,  a
priori, these systems may contain  a WD and a chromospherically active
main       sequence       companion,        e.g.        V471       Tau
\citep{hussainetal06-1}\footnote{$Simbad$  lists 2,  16 and  6 of  our
  objects    as    quasars,    galaxies   and    globular    clusters,
  respectively. However, we do not exclude these sources from our list
  as visual inspection  of their LAMOST spectra  clearly revealed they
  are FGK main  sequence stars.}. This left us with  a total number of
2,168  spectra of  1,549 unique  WD+FGK binary  candidates. This  list
includes  the  504  LAMOST  DR\,1  objects  we  already  presented  in
\citet{parsonsetal16-1} except  for 13  sources that were  excluded by
LAMOST due to pipeline updates.  Table \ref{tab:wdfgk-sample} includes
all the 1,549  objects and the corresponding  duplicate spectra, where
we also provide  coordinates, stellar parameters as  obtained from the
DR\,4  stellar parameters  pipeline, $GALEX$  and optical  magnitudes,
spectral  ID identifiers,  signal-to-noise  ratio of  the spectra  and
$Simbad$ classification  when available.  In the  following section we
present  a  follow-up radial  velocity  survey  for identifying  close
binaries among the 1,549 WD+FGK systems.

\begin{table}
\centering
\caption{\label{t-log} Log of the  observations.  We include here only
  follow-up spectroscopy obtained  by our team. The MJD  of all LAMOST
  DR\,4   spectra    used   in   this    work   can   be    found   in
  Table\,\ref{tab:wdfgk-sample}.  We  indicate the telescopes  and the
  spectrographs used, the resolving powers of the spectra, the type of
  observation (visitor or service), the date and the number of spectra
  obtained per night.} \setlength{\tabcolsep}{1.5pt}
\begin{tabular}{cccccc}
\hline
\hline
 Telesc.    & Spectrograph & R      & Type    & Date              & \#Spectra   \\
 \hline
Palomar 5.1 & DBSP         & 6,370   & Visitor & 2014 Jan 19       & 36 \\
Palomar 5.1 & DBSP         & 6,370   & Visitor & 2014 Jan 21       & 9  \\
CFHT        & Espadons     & 68,000  & Service & 2015 Aug--        & 48 \\
CFHT        & Espadons     & 68,000  & Service & --2016 Jan        &    \\
TNT         & MRES         & 15,000  & Visitor & 2015 Feb 3        & 5  \\ 
TNT         & MRES         & 15,000  & Visitor & 2015 Feb 4        & 11 \\
TNT         & MRES         & 15,000  & Visitor & 2015 Feb 5        & 2  \\
TNT         & MRES         & 15,000  & Visitor & 2015 Feb 6        & 1  \\ 
TNT         & MRES         & 15,000  & Visitor & 2015 Apr 1        & 7  \\ 
TNT         & MRES         & 15,000  & Visitor & 2015 Apr 2        & 6  \\ 
TNT         & MRES         & 15,000  & Visitor & 2015 May 11       & 2  \\
TNT         & MRES         & 15,000  & Visitor & 2015 May 12       & 4  \\
TNT         & MRES         & 15,000  & Visitor & 2015 May 13       & 1  \\
TNT         & MRES         & 15,000  & Visitor & 2016 Apr 13       & 7  \\
TNT         & MRES         & 15,000  & Visitor & 2016 Apr 27       & 1  \\
TNT         & MRES         & 15,000  & Visitor & 2016 Apr 28       & 3  \\
TNT         & MRES         & 15,000  & Visitor & 2017 Mar 4        & 6  \\
TNT         & MRES         & 15,000  & Visitor & 2017 Mar 5        & 9  \\
TNT         & MRES         & 15,000  & Visitor & 2017 Mar 6        & 11 \\
TNT         & MRES         & 15,000  & Visitor & 2017 Apr 27       & 3  \\
TNT         & MRES         & 15,000  & Visitor & 2017 Apr 29       & 1  \\
TNT         & MRES         & 15,000  & Visitor & 2017 Apr 30       & 4  \\
INT         & IDS          & 6,600   & Visitor & 2016 Jan 23       & 15 \\
INT         & IDS          & 6,600   & Visitor & 2016 Jan 24       & 19 \\
INT         & IDS          & 6,600   & Visitor & 2016 Jan 25       & 26 \\
INT         & IDS          & 6,600   & Visitor & 2016 Nov 15       & 1  \\
INT         & IDS          & 6,600   & Visitor & 2016 Nov 16       & 20 \\
INT         & IDS          & 6,600   & Visitor & 2016 Nov 19       & 12 \\
INT         & IDS          & 6,600   & Visitor & 2016 Nov 20       & 8  \\
INT         & IDS          & 6,600   & Visitor & 2016 Nov 21       & 15 \\
INT         & IDS          & 6,600   & Service & 2016 Nov 18       & 9  \\ 
INT         & IDS          & 6,600   & Service & 2016 Nov 19       & 26 \\
SPM 2.1     & Echelle      & 20,000  & Visitor & 2015 Dec 5        & 9  \\
SPM 2.1     & Echelle      & 20,000  & Visitor & 2015 Dec 6        & 18 \\
SPM 2.1     & Echelle      & 20,000  & Visitor & 2015 Dec 7        & 12 \\
TNG         & HARPS-N      & 115,000 & Visitor & 2017 Jan 1        & 25 \\
TNG         & HARPS-N      & 115,000 & Visitor & 2017 Jan 2        & 25 \\
TNG         & HARPS-N      & 115,000 & Visitor & 2017 Jan 3        & 24 \\
TNG         & HARPS-N      & 115,000 & Visitor & 2017 Jan 4        & 23 \\
\hline
\end{tabular}
\end{table}

\section{Observations}
\label{s-obs}

In the previous  section we have provided details on  our strategy for
efficiently identifying  WD+FGK binaries. However, it  is important to
keep in  mind that only  a fraction  of these systems  ($\sim$1/4; see
e.g. \citealt{willems+kolb04-1}) are expected  to have evolved through
mass-transfer interactions such as a  common envelope phase and become
close compact binaries of orbital periods  ranging from a few hours to
several weeks.  In  order to identify these close  binaries we require
radial  velocity information,  which  we obtained  from the  available
LAMOST DR\,4  spectra as well  as from our own  follow-up spectroscopy
taken at several  different telescopes around the globe (see  a log of
the observations in Table\,\ref{t-log}).  In the following we describe
the spectroscopic observations.

\begin{figure}
  \begin{center}
    \includegraphics[width=\columnwidth]{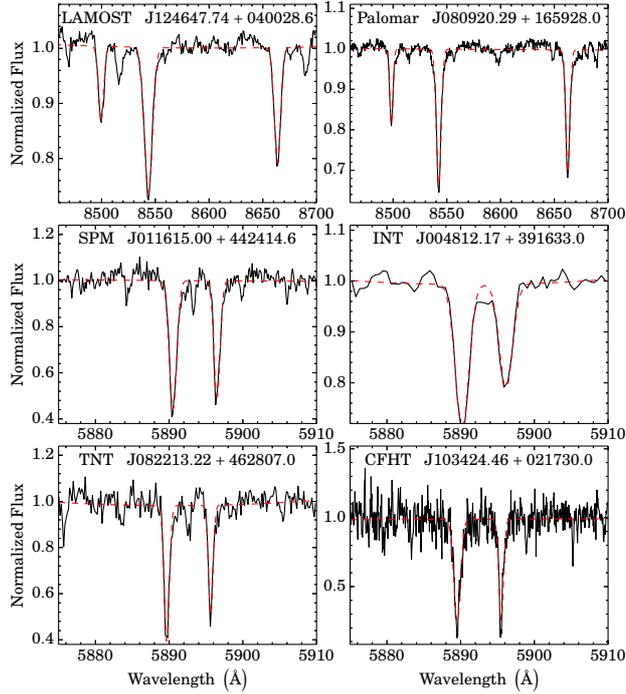}
    \caption{\label{f-rvs} Fits (red dashed lines) to the \Ion{Ca}{II}
      triplet  at  $\sim$8,500\AA\  and  the  \Ion{Na}{I}  doublet  at
      $\sim$5,800\AA\  (black solid  lines) for  measuring the  radial
      velocities.  The telescopes used  and object names are indicated
      in each panel.}
  \end{center}
\end{figure}

\subsection{LAMOST DR\,4}

As mentioned above, the current data product of LAMOST is DR\,4, which
includes 4\,202\,127  spectra of  late A and  FGK main  sequence stars
with available stellar parameters. Among  these we have selected 1,549
as WD+FGK  binary candidates (Section\,\ref{s-selection}).   The total
number of combined  available LAMOST spectra for  our selected targets
is  2,168,   or  5,357  if   we  take  into  account   the  individual
sub-exposures of each combined spectrum.  We also cross-correlated our
list of LAMOST WD+FGK binaries with the spectroscopic data base of the
Sloan Digital Sky Survey \cite[SDSS;][]{yorketal00-1} data release 12,
which resulted  in 88 additional  spectra for  21 of our  targets. The
resolving power of both LAMOST and SDSS spectra is $R\sim$1,800.

\subsection{Palomar Hale Telescope}
\label{s-palomar}

Follow-up spectroscopy for 44 WD+FGK  binaries was obtained during two
nights from the  5.1-meter Palomar Hale Telescope in  2014, January 19
and 21 (Table\,\ref{t-log}) at Palomar  Observatory in north San Diego
County, California, USA.  Two different  gratings were used, the 1,200
lines\,mm$^{-1}$ and  600 lines\,mm$^{-1}$,  providing spectra  of the
red  and blue  arms respectively.   For this  the double  spectrograph
(DBSP),  together with  a  1-arcsec width  long-slit  were used.   Arc
spectra were taken at the beginning of  the nights and we used the sky
lines to account the flexure of the instrument (note that we only made
use of  the red arm  spectra to  measure the radial  velocities).  The
resolving   power   was   6,370   in  the   red   arm   covering   the
$\sim$7,600-9,300\,\AA\,  wavelength  range.   The   blue  arm  had  a
resolving power of 1,380, ranging $\sim$3,500-6,500\,\AA.

\subsection{Isaac Newton Telescope}

Eight nights  in December  2014, January 2016  and November  2016 plus
four   hours   of  service   mode   observations   in  November   2016
(Table\,\ref{t-log}) at the the Iasaac Newton Telescope (INT) in Roque
de los  Muchachos Observatory  at La  Palma, Spain  provided follow-up
spectroscopy for 47  WD+FGK binaries.  The spectra  were obtained from
the 1200Y  grating.  A 1-arcsec  wide long-slit together with  the IDS
(Intermediate Dispersion Spectrographs) were used throughout the whole
observations.    The   spectra  covered   the   $\sim$5,100-6,400\AA\,
wavelength range at a resolving power  of 6,600.  One arc spectrum was
taken at  the beginning each  night.  We calibrated the  spectra using
the single arc spectrum and adjusted their calibration using the night
sky emission lines.

\subsection{Canada-France-Hawaii Telescope}

The   Canada-France-Hawaii   Telescope   (CFHT)   provided   follow-up
spectroscopy for 25 WD+FGK binaries.  The observations were carried in
service mode from August 2015  to January 2016 (Table\,\ref{t-log}) at
Mauna Kea  Observatory, Hawaii,  USA. For this  a 1.6  arcsec aperture
hole was used together  with the Espadons (Echelle SpectroPolarimetric
Device  for  the  Observation  of Stars)  spectrograph.   The  spectra
covered the  entire optical wavelength  range at a resolving  power of
68\,000.  Arc spectra were obtained before each science spectrum.

\begin{figure}
  \begin{center}
    \includegraphics[angle=-90,width=\columnwidth]{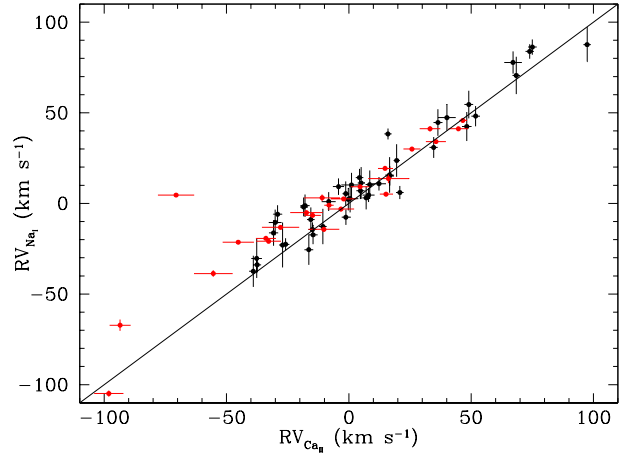}
    \caption{\label{f-rvcheck}  Comparison  of the  radial  velocities
      measured   from   the    \Ion{Na}{I}   absorption   doublet   at
      $\sim$5,800\AA\, and from the \Ion{Ca}{II} absorption triplet at
      $\sim$8,500\AA\, of  the same low-resolution LAMOST  (black) and
      high-resolution CFHT (red) spectra of 45 and 24 WD+FGK binaries,
      respectively.}
  \end{center}
\end{figure}

\subsection{Thai National Telescope}

Additional follow-up spectroscopy for 30 WD+FGK binaries were obtained
from  the  Thai   National  Telescope  (TNT)  at   the  That  National
Observatory in Thailand during 20  nights spread between 2015 and 2017
(see Table\,\ref{t-log}).   Using a 1.4-arscsec slit  width, the light
was redirected to the Middle Resolution fiber-fed Echelle Spectrograph
(MRES).  The resulting spectra covered the 3,900-8,800\AA\, wavelength
range at a  resolving power of 15,000.  Arc spectra  were taken at the
beginning of each night  and we used the sky lines  to account for the
flexure of the spectrograph.

\subsection{San Pedro M\'artir 2.12m Telescope}

Follow-up spectroscopy of 5 WD+FGK  binaries was obtained during a run
in 2015 December 5, 6 and 7 (see Table\,\ref{t-log}) using the Echelle
spectrograph attached to  the 2.12\,m telescope at  San Pedro M\'artir
Observatory in Baja California, M\'exico.  We used a slit width of 2.8
arcsec,  resulting in  3,650 --  7,300\AA\@ wavelength  coverage at  a
resolving power R $\sim$ 20,000.  Arc spectra were obtained before and
after each object.

\begin{table}
\centering
\caption{The  radial  velocities of  the  1,453  LAMOST WD+FGK  binary
  candidates.  The Heliocentric Julian Dates (HJD) and telescopes used
  for obtaining the spectra are  also listed.  ``-9999" indicates that
  no radial velocity is available.  The complete table can be found in
  the electronic version of the paper.}
\label{tab:rv_subexp}
\setlength{\tabcolsep}{1.5pt}
\begin{tabular}{ccccc}
\hline
  Name   &       HJD     &       RV      &       Err     &       Telescope\\
         &       (days)   & (\kms) & (\kms) &  \\
\hline
J000230.57+155618.1     &       2456199.15371   &       -24.45  &       11.64   &       LAMOST  \\
J000230.57+155618.1     &       2456199.16547   &       -16.71  &       13.52   &       LAMOST  \\
J000232.64+074305.5     &       2457309.07594   &       -9999   &       -9999   &       LAMOST  \\
J000232.64+074305.5     &       2457309.09912   &       -9999   &       -9999   &       LAMOST  \\
J000232.64+074305.5     &       2457309.12228   &       -9999   &       -9999   &       LAMOST  \\
J000243.58+270214.7     &       2456647.98403   &       -23.47  &       11.62   &       LAMOST  \\
J000243.58+270214.7     &       2456647.99553   &       -29.21  &       10.97   &       LAMOST  \\
J000243.58+270214.7     &       2456648.00691   &       -23.53  &       10.70   &       LAMOST  \\
J000324.28+063732.3     &       2456618.03752   &        88.30  &       10.26   &       LAMOST  \\
J000324.28+063732.3     &       2456618.04911   &        85.30  &       10.29   &       LAMOST  \\
\hline
\end{tabular}
\end{table}

\subsection{Telescopio Nazionale Galileo}

Four  nights  of  observations  were carried  out  at  the  Telescopio
Nazionale  Galileo in  June 2017  (see Table\,\ref{t-log}),  providing
follow-up spectroscopy of 45 additional  WD+FGK binaries.  We used the
HARPS-N  (High  Accuracy Radial  velocity  Planet  Searcher --  North)
spectrograph  equipped  and  the  1\,arcsec fibre,  resulting  in  the
wavelength coverage of $\sim$3,800-6,900\,\AA.  The resolving power of
HARPS-N is 115000.  We used a  second fibre to obtain arc lamp spectra
that were then used to establish a generic pixel-wavelength relation.

\section{Radial Velocities}
\label{s-rvs}

Including  all  LAMOST  DR\,4,  SDSS  DR\,12  and  our  own  follow-up
spectroscopy we have at hand a total of 5,885 spectra for 1,453 WD+FGK
binary candidates.  We made use of these spectra to measure the radial
velocities of  our WD+FGK binaries  as follows (the  radial velocities
are provided in Table\,\ref{tab:rv_subexp}).

We followed the  method described by \citet{rebassa-mansergasetal08-1}
and   \citet{renetal13-1}   to   fit  the   \Ion{Na}{I}   doublet   at
$\sim$5,800\AA\, arising from all spectra  obtained from our INT, SPM,
CFHT and TNT observations (Table\,\ref{t-log}).  To that end we used a
combination  of  a  second  order  polynomial  and  a  double-Gaussian
profile.

\begin{figure*}
  \begin{center}
    \includegraphics[angle=-90,width=\textwidth]{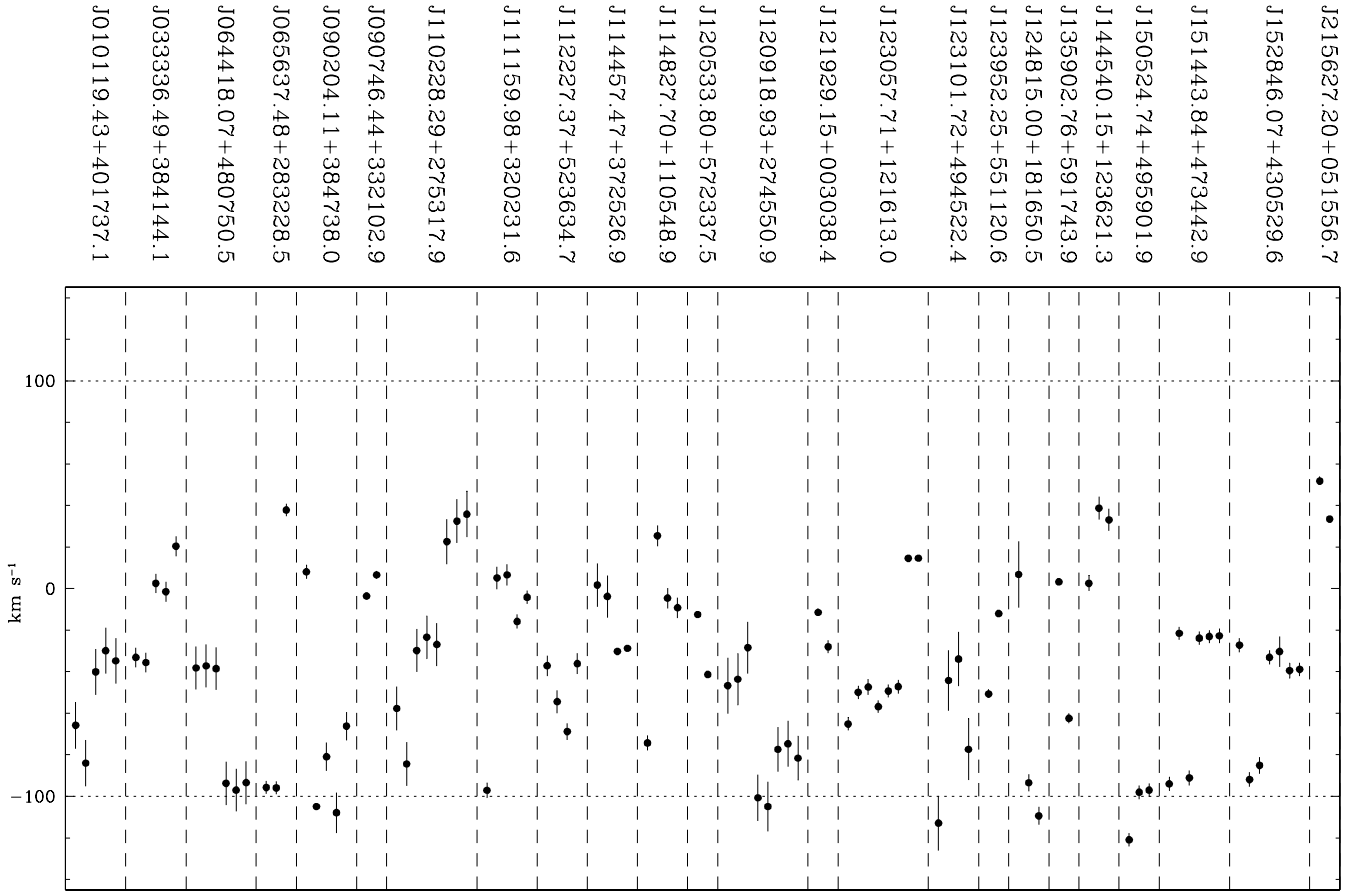}
    \caption{\label{f-rvs} The  radial velocities measured for  the 24
      WD+FGK binaries  that display  significant ($>3  \sigma$) radial
      velocity variation.  The object names are provided in the top.}
  \end{center}
\end{figure*}

We used the same approach as  above described for measuring the radial
velocities from the  LAMOST and SDSS spectra.  However,  we noted that
in most cases the \Ion{Na}{I}  doublet profile was poorly sampled thus
resulting in radial velocities  associated to large uncertainties.  We
hence searched for common, clean and sufficiently resolved atmospheric
features typical of F, G and K stars (we remind the reader that the WD
features are  not visible in  the optical spectra of  WD+FGK binaries)
and found  that the \Ion{Ca}{II}  triplet at $\sim$8,500\AA\,  was the
most frequently  visible, cleanest  and strongest.  We  hence measured
the  LAMOST  and  SDSS  radial  velocities  fitting  the  \Ion{Ca}{II}
absorption triplet with a combination of a second order polynomial and
a  triple-Gaussian profile  of fixed  separations.  We  used the  same
approach to measure the radial  velocities from the red-arm spectra of
our Palomar  data since the  blue-arm spectra covering  other features
such  as   the  \Ion{Na}{I}  doublet  at   $\sim$5,800\AA\,  were  not
sufficiently resolved due  to the low resolving  power ($R=1,380$, see
Section\,\ref{s-palomar}).   A  few  examples  of  double  and  triple
Gaussian fits to  the \Ion{Ca}{II} and \Ion{Na}{I}  profiles are shown
in Figure\,\ref{f-rvs}.

Finally,  we used  the  HARPS-N  pipeline \citep{lovis+pepe07-1}  that
makes  use of  the  cross-correlation  function \citep{queloz95-1}  to
derive the  radial velocities from  the spectra obtained from  our TNG
observations.

\begin{figure*}
  \begin{center}
    \includegraphics[angle=-90, width=\columnwidth]{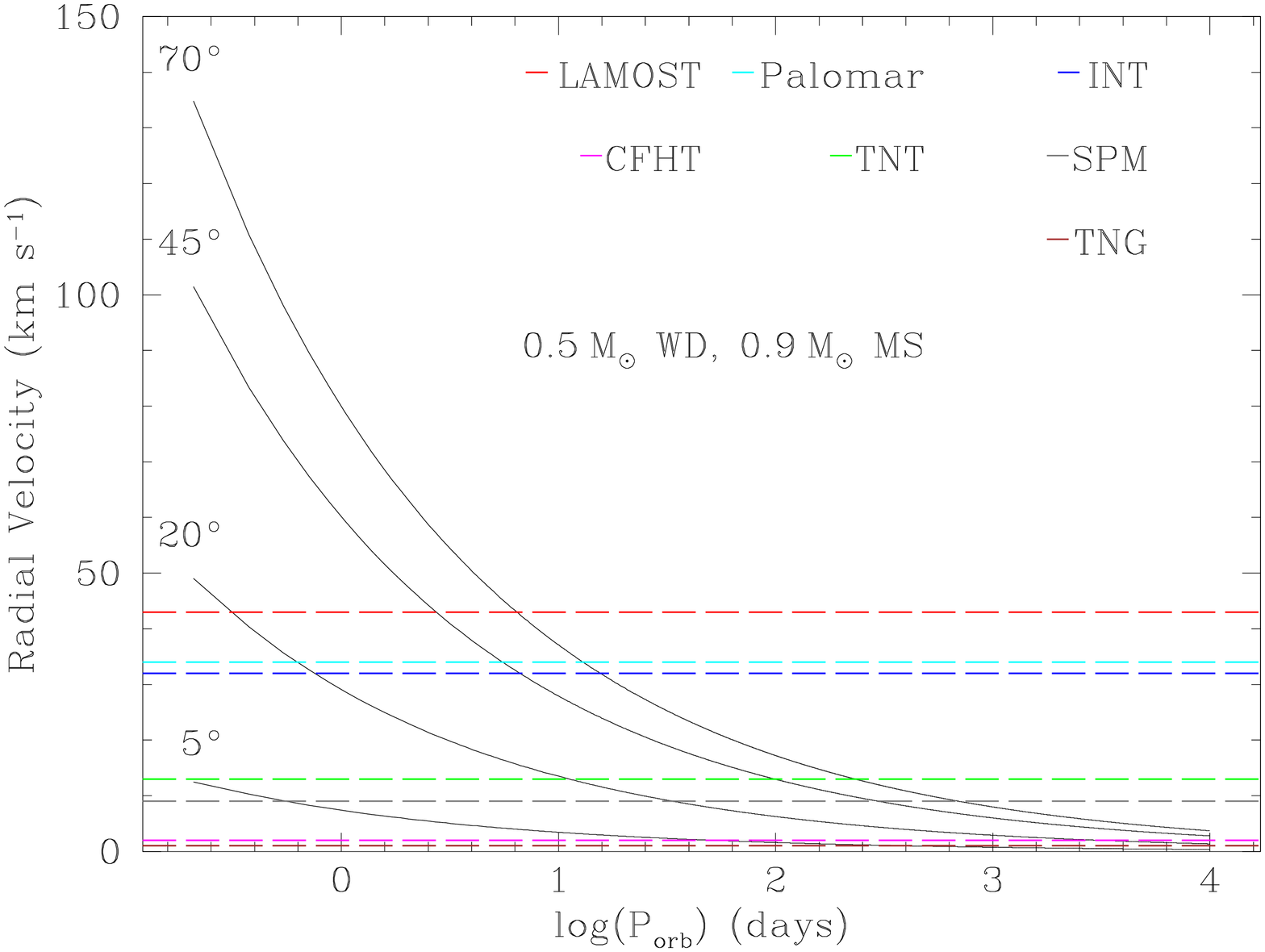}
    \includegraphics[angle=-90, width=\columnwidth]{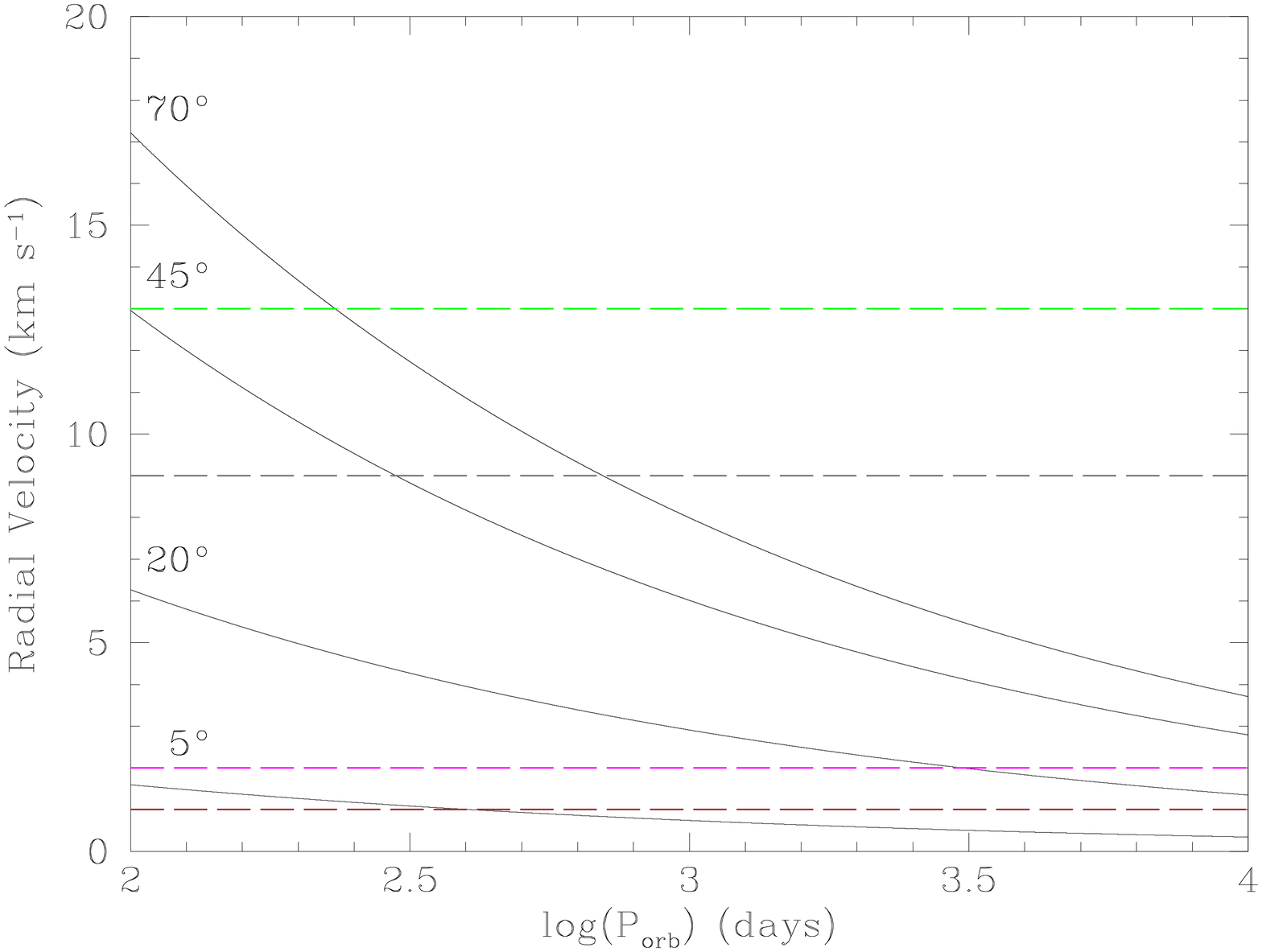}
    \includegraphics[angle=-90, width=\columnwidth]{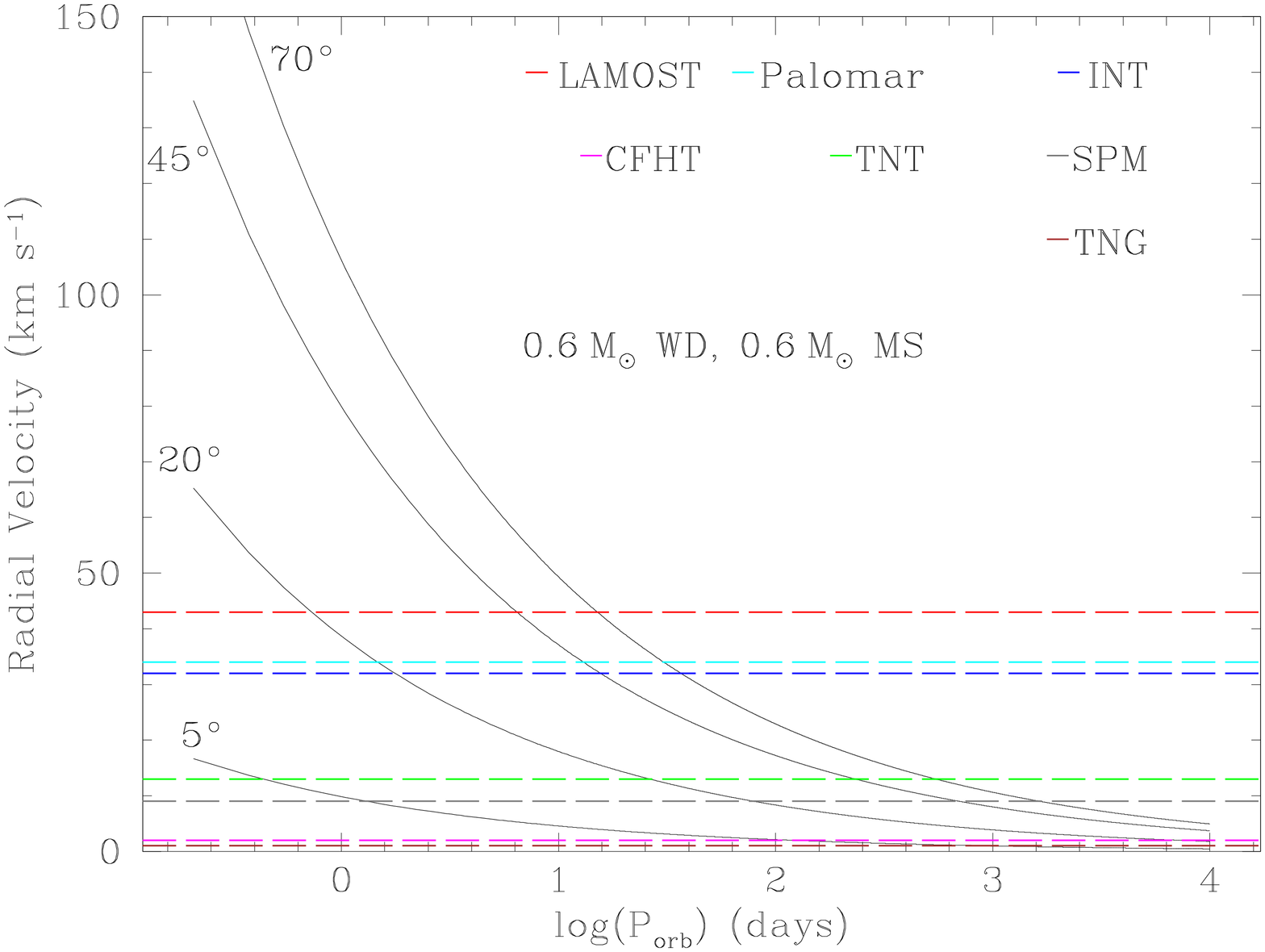}
    \includegraphics[angle=-90, width=\columnwidth]{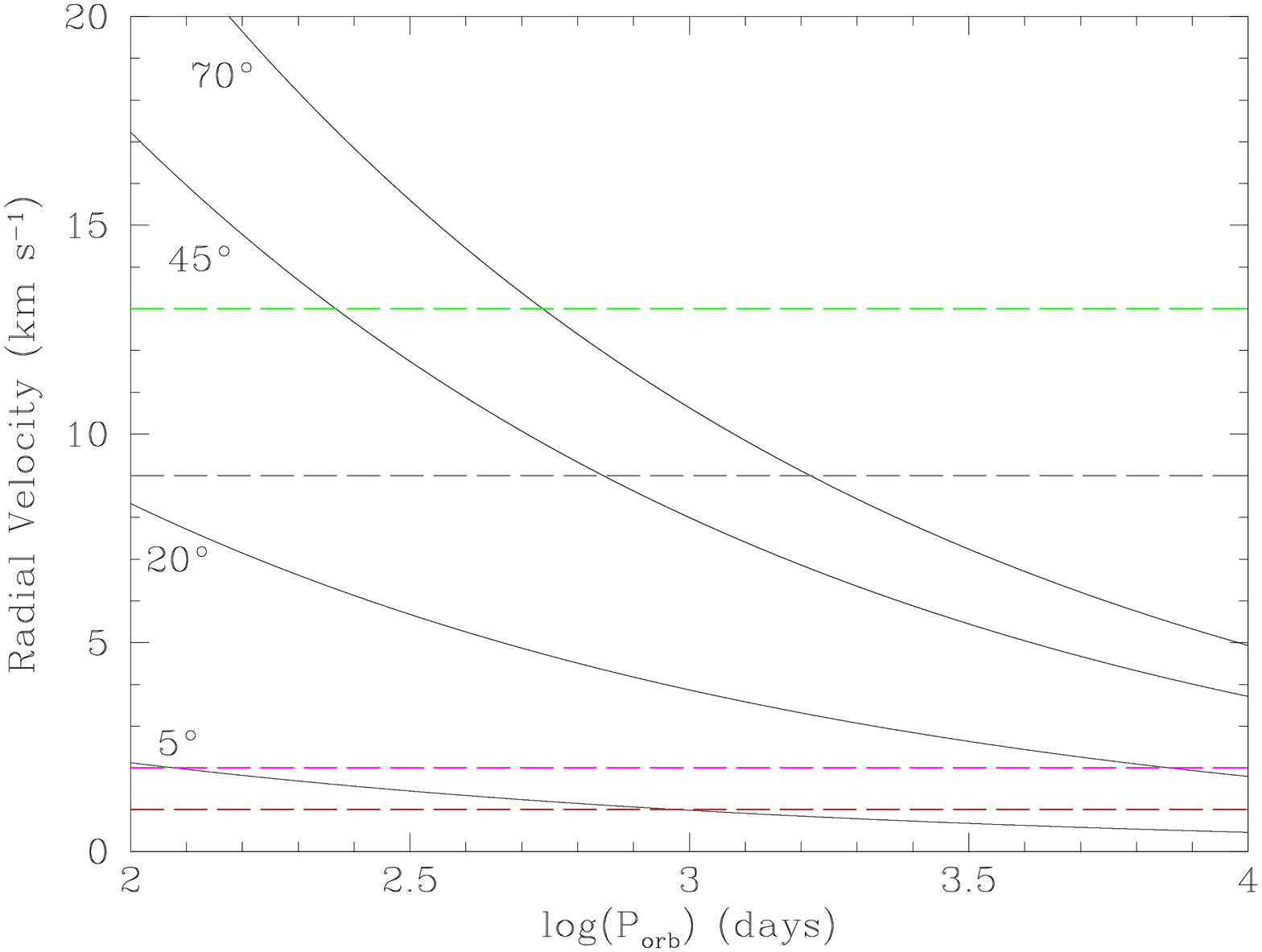}
    \includegraphics[angle=-90, width=\columnwidth]{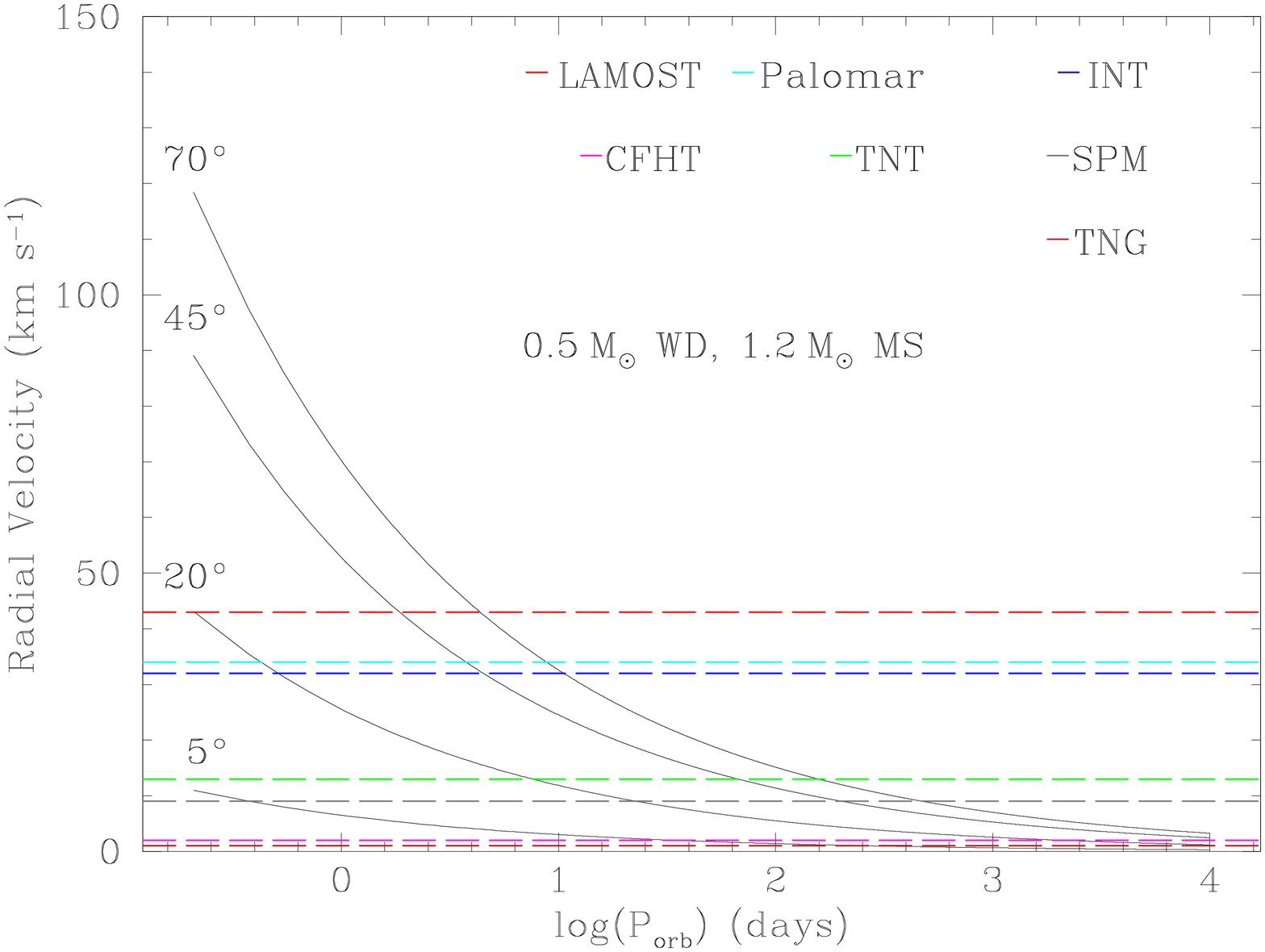}
    \includegraphics[angle=-90, width=\columnwidth]{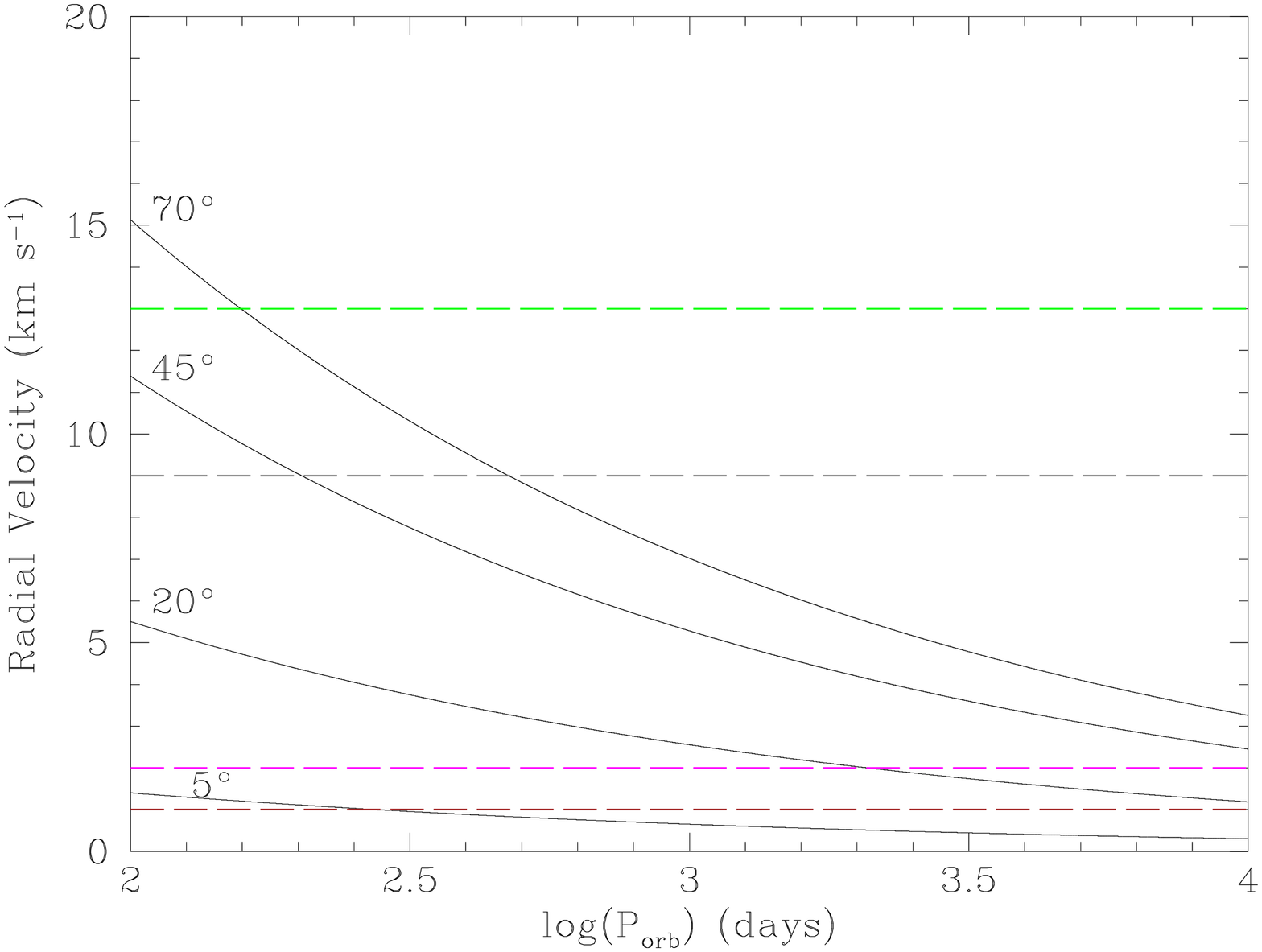}
    \caption{\label{f-thres}  In black  solid lines  we represent  the
      semi-amplitude radial  velocities of  the main sequence  stars in
      WD+FGK binaries formed  by a 0.5\,\Msun\, WD  and a 0.9\,\Msun\,
      main  sequence  star  (top  panels), a  0.6\,\Msun\,  WD  and  a
      0.6\,\Msun\,   main  sequence   star  (middle   panels)  and   a
      0.5\,\Msun\, WD and a 1.2\,\Msun\, main sequence (bottom panels)
      as a function of orbital  period for the assumed inclinations of
      5, 20, 45 and 70 degrees. The  right panels are simply a zoom in
      to the longer orbital periods considered.  The horizontal dashed
      lines indicate the 3$\sigma$  radial velocity variation that the
      telescopes/instruments  used  in  this   work  can  detect.   As
      expected, the 3$\sigma$ radial  velocity variation decreases for
      increasing values of the spectral resolution.}
  \end{center}
\end{figure*}

\section{Identification of close WD+FGK binaries}

We identified  close WD+FGK  binaries via  a radial  velocity analysis
based on multi-epoch spectroscopy. That  is, we considered a system to
be  a close  binary if  we detected  significant ($>3  \sigma$) radial
velocity  variation.  To  that end  we combined  all available  radial
velocities for  each object,  independently on  the resolution  of the
spectra from which they  were measured\footnote{The timescale in which
  we detect radial  velocity variations varies from  system to system,
  ranging  from one  day to  $\sim$3 years.   The reason  for this  is
  twofold.   Firstly,  observations  were  done in  both  visitor  and
  service mode  (with service  mode observations obviously  spanning a
  wider time  range), and secondly  we observed some systems  with the
  same telescope (e.g.,  TNT and INT) in  different years.}.  However,
it is important to keep in mind that we measured the radial velocities
adopting   three    different   methods   depending   on    the   data
(Section\,\ref{s-rvs}),  i.e.   fitting  the \Ion{Ca}{II}  triplet  at
$\sim$8,500\AA, fitting the \Ion{Na}{I} doublet at $\sim$5,800\AA\, or
applying        the       cross-correlation        function       (see
Section\,\ref{s-rvs}). Whilst both  the cross-correlation function and
the \Ion{Ca}{II} triplet  fitting technique have been  proven to yield
precise    radial   velocities    --    see   \cite{queloz95-1}    and
\cite{kunderetal17-1}, respectively -- the \Ion{Na}{I} doublet profile
at   $\sim$5,800   may   be  affected   by   interstellar   absorption
\citep{welshetal10-1}.  In  order to  explore possible effects  of the
interstellar  absorption  on  the measurement  of  \Ion{Na}{I}  radial
velocities,  we compare  in  Figure\,\ref{f-rvcheck} the  \Ion{Ca}{II}
absorption  triplet  and  the \Ion{Na}{I}  absorption  doublet  radial
velocities measured from 45  LAMOST high signal-to-noise ratio spectra
($>$300)\footnote{The  \Ion{Na}{I} doublet  is poorly  sampled by  the
  low-resolution LAMOST spectra, hence we require high signal-to-noise
  ratio for reducing the radial  velocity uncertainties.}  and 24 CFHT
spectra  of unique  objects.  It  becomes  obvious by  looking at  the
figure that  the radial velocities  are in relatively  good agreement,
although discrepancies arise for some  systems.  Hence, in cases where
a given  WD+FGK binary has  available radial velocities  measured from
both  line profiles,  we decide  to not  to combine  them in  order to
search  for radial  velocity variations.   Moreover, we  only consider
those values with uncertainties  below 20~\kms.  This exercise results
in 24 systems displaying more than 3$\sigma$ radial velocity variation
that we classify  as close binaries.  The radial velocities  of the 24
objects are  displayed in  Figure\,\ref{f-rvs}, where we  also include
the objects names.  It is worth  mentioning that one of our identified
close  binaries  is  classified  as  an  RS  CVn  binary  by  $Simbad$
(J033336.49+384144.1,   with  an   orbital   period   of  2.85   days;
\citealt{drakeetal14-1}).

\section{Discussion}

In the  previous section we  identified 24 WD+FGK  binaries displaying
significant radial velocity variations. The  number of systems with at
least  two available  radial  velocities that  do  not display  radial
velocity  variation is  1,402, which,  at face  value, yields  a close
binary  fraction  of  $\sim$1.5  per  cent.  However,  this  value  is
misleading due to the following reasons.

First,  in  $\sim$70  per  cent  of the  cases  the  available  radial
velocities  are  obtained  during  the  same  night,  which  decreases
considerably the probability of  detecting close binaries with orbital
periods longer than one day.  If  we consider only WD+FGK systems with
available radial  velocities measured from spectra  taken separated by
at least  one night,  then the  number of  systems displaying  and not
displaying more than  3$\sigma$ radial velocity variations  are 24 and
395 respectively, i.e.  a revised close binary fraction of $\sim$6 per
cent.

Second, low-inclination systems and  systems with long orbital periods
are  harder to  identify.   This  is not  only  because the  detection
probability  decreases (see  e.g.   \citealt{nebotetal11-1}) but  also
because the radial  velocity amplitudes of such  systems are generally
lower, hence  higher resolution  spectra are  needed to  derive radial
velocities  with  lower  uncertainties suitable  for  detecting  small
variations. This effect is important here since we used a wide variety
of telescopes and  instruments of resolving powers  ranging from 1,800
(LAMOST) to 115,000  (TNG/HARPS). In order to quantify  this effect we
consider a binary formed  by a WD of 0.5\Msun\ --  a typical value for
WDs in close WD plus main  sequence binaries that evolved through mass
transfer  episodes,  see  \cite{rebassa-mansergasetal11-1}  --  and  a
secondary star  mass of  0.9\Msun\, (which  corresponds to  a spectral
type of $\sim$G5,  the median value in our  sample).  Assuming orbital
inclinations  of   5,  20,  45   and  70  degrees  we   calculate  the
semi-amplitude velocities of  the  main sequence  stars employing  the
third law  of Kepler for orbital  periods up to 10,000  days.  This is
illustrated in  the top  panels of Figure\,\ref{f-thres}  (black solid
curves).  We  then consider  the 3$\sigma$ radial  velocity variations
that each  combination of  telescope/instrument used  in this  work is
able to detect.  We do this assuming a radial velocity error of 10 per
cent the spectral resolution.  The results are shown in the top panels
of Figure\,\ref{f-thres}  as dashed horizontal lines.   Inspecting the
figures it can be clearly seen that the LAMOST data are inefficient at
detecting close  binaries with orbital inclinations  below 20 degrees,
but that close binaries with higher inclinations should be detected up
to orbital  periods of $\sim$3 days  for an inclination of  45 degrees
and $\sim$6 days  for an inclination of 70 degrees  (assuming that the
spectra are obtained sufficiently  separated to avoid sampling similar
orbital  phases,  see  next  paragraph).  The  orbital  period  limits
increase for the INT  and Palomar data, where we expect  to be able to
identify  all  close  WD+FGK  binaries  with  orbital  periods  up  to
$\sim$8.5 days and orbital inclinations  higher than 70 degrees.  From
our higher resolution spectra we expect  be able to identify all close
WD+FGK binaries  with orbital periods  up to  100 days (CFHT)  and 300
days (TNG) at orbital inclinations $\ga$5 degrees (again assuming that
the spectra  are taking at  different orbital phases).   These orbital
period limits rely entirely on the assumption of a 0.5\Msun\, WD and a
0.9\Msun\, main sequence star (the most likely binary pair to be found
among  our  sample of  WD+FGK  binaries),  but  of course  many  other
combinations  are  possible.   For  completeness,  we  show  two  more
examples  in  Figure\,\ref{f-thres},  middle   (0.6\Msun\,  WD  and  a
0.6\Msun\,  main  sequence  star)  and bottom  (0.5\Msun\,  WD  and  a
1.2\Msun\, main sequence star) panels.  One can see that, although the
orbital period  limits change  from one case  to another,  the overall
results  are unchanged,  i.e.   the higher  resolution  data are  more
suitable  to detect  longer  orbital period  and/or lower  inclination
systems.   Thus, if  we only  take  into account  WD+FGK systems  with
available  radial velocities  separated  by more  than  one night  and
observed by the CFHT and  TNG telescopes (our higher resolution data),
the number  of systems displaying  and not displaying  radial velocity
variations is 6 and 58, i.e.   a close binary fraction of $\sim$10 per
cent.

Third, we need to take into account that the radial velocity values of
a given system  may be taken at the same  orbital phase, which implies
that  some  objects we  catalogue  as  WD+FGK  binaries that  did  not
interact in  their past are in  fact close binaries.  Moreover,  it is
important to keep in mind that, although our methodology for selecting
WD+FGK  systems  based   on  $GALEX$  colours  is   efficient  --  see
Section\,\ref{s-selection}  --  the  probability exists  that  we  are
selecting other kind of astronomical objects such as e.g.  active main
sequence     stars      \citep{smithetal14-1}     or      RR     Lyrae
(Section\,\ref{s-selection}). Quantifying the contamination from other
sources in our sample requires UV spectra of a large number of sources
to visually  test the  presence of  the WD. The  fact that  we clearly
observed the WD features in the Hubble Space Telescope of nine objects
in our sample seems to indicate that the contamination should be small
\citep{parsonsetal16-1}.

Because of  these reasons, the  close binary fraction of  $\sim$10 per
cent should be taken as preliminary.  This value is considerably lower
than the the close binary fraction  resulting from our analysis of WDs
in orbits with M dwarf  companions (WD+M binaries) \citep[$\sim$25 per
  cent][]  {nebotetal11-1}. This  is  likely because  the WD+M  sample
contains fully convective stars for which magnetic braking is expected
to  be   largely  reduced   \citep{reiners+basri09-1,  morinetal10-1}.
Hence,  angular momentum  loss is  much more  inefficient at  bringing
these  (WD+M)  binary  components  together  for  initiating  a  semi-
detached    phase    \citep{politano+weiler06-1,    schreiberetal10-1,
  rebassa-mansergasetal13-1,  zorotovicetal16-1}. In  other words,  we
expect  a   larger  fraction  of   WD+FGK  binaries  to   have  become
semi-detached  than WD+M  systems,  in line  with  the observed  close
binary fractions. An  additional possibility may be  that close WD+FGK
systems emerge  from common envelope at  longer periods (weeks/months)
than WD+M  systems, hence they  are harder to identify  (especially by
our lower  resolution data).  This would  imply a  higher common
  envelope efficiency for  the WD+FGK systems.  Indeed,  this seems to
  be the case for the  two long-orbital period WD+FGK binaries studied
  by \citet{zorotovicetal14-1}.

\section{Conclusions}

With  the   aim  of  testing   the  importance  of  the   single-  and
double-degenerate  channels for  Type Ia  supernovae (SN\,Ia)  we have
started  an observing  campaign dedicated  to identify  detached close
WD+FGK binaries  and to  measure their  orbital periods  and component
masses.   With these  data at  hand we  will be  able to  predict what
systems will initiate  a phase of thermal-timescale  mass transfer and
thus become super-soft sources  and potential single-degenerate SN\,Ia
progenitors, and  what binaries  will evolve  through a  second common
envelope  phase   and  form  double  white   dwarfs,  i.e.   potential
double-degenerate SN\,Ia progenitors.  In this  paper, the second of a
series of  publications, we identified 1,549  WD+FGK binary candidates
from the data release 4 of LAMOST and presented follow-up spectroscopy
of 1,453 of them.  Analysing the radial velocities we have detected 24
close binaries  displaying significant  radial velocity  variation and
calculated a preliminary close binary fraction among WD+FGK systems of
$\sim$10 per cent. We have also shown that high-resolution spectra are
needed  to efficiently  identify  lower  inclination ($\la$5  degrees)
and/or longer  orbital period systems  ($\ga$100 days; in  these cases
service  mode  observations  are   additionally  required).   This  is
important  for detecting  WD+FGK systems  that will  evolve through  a
second  common  envelope  phase  and  avoid  merging  to  thus  become
double-degenerate  SN\,Ia progenitors  candidates.   In a  forthcoming
publication  we  will  present additional  follow-up  observations  of
WD+FGK  binaries  identified  by  the  RAVE  survey  in  the  southern
hemisphere as well as our first orbital periods measured.

\section*{Acknowledgments}

This  research  has  been  funded  by  the  AGAUR,  by  MINECO  grants
AYA2014-59084-P  (ARM, EGB)  and  AYA2014-55840-P (EV),  by the  Young
Researcher  Grant  of  National  Astronomical  Observatories,  Chinese
Academy of Sciences (JJR), by the  Key Basic Research Program of China
(XL; 2014CB845700),  by the  Leverhulme Trust  (SGP), by  the European
Research  Council   under  the  European  Union's   Seventh  Framework
Programme (BTG; FP/2007-2013)/ERC  Grant Agreement n.320964 (WDTracer)
and  by  Milenium Science  Initiative,  Chilean  Ministry of  Economy,
Nucleus P10-022-F (MRS).

This work has made use of data products from the Guoshoujing Telescope
(the  Large  Sky  Area  Multi-Object  Fibre  Spectroscopic  Telescope,
LAMOST). LAMOST  is a National  Major Scientific Project built  by the
Chinese Academy of Sciences. Funding for the project has been provided
by the National Development and  Reform Commission. LAMOST is operated
and  managed  by  the  National  Astronomical  Observatories,  Chinese
Academy of Sciences.

Based on  observations performed at  the Palomar Hale  Telescope (time
obtained  through Chinese  Telescope  Access  Program, 2013B-20),  the
Isaac  Newton Telescope  (programs 2014B-80,  2015B-7, 2016B-33),  the
Canada  France   Hawaii  Telescope  (time  obtained   through  Chinese
Telescope Access Program, 2015B-27),  the Thai National Telescope, the
2.12~m telescope in San Pedro  M\'artir Observatory and the Telescopio
Nazionale Galileo (program CAT17A-5).

\label{lastpage}

\end{document}